\documentclass[reprint,twocolumn,aps,prl,showpacs,longbibliography]{revtex4-2}

\usepackage{amsmath}
\usepackage{amssymb}
\usepackage{graphicx}
\usepackage{dcolumn}
\usepackage{bm}
\usepackage[colorlinks=true, allcolors=blue]{hyperref}
\usepackage{epstopdf}
\usepackage[utf8]{inputenc}
\usepackage[dvipsnames]{xcolor}
\usepackage{appendix}

\usepackage{enumerate}

\begin{document}

\title{Local Spin Excitations Mediate Quasiparticle Breakdown in the Orbital-Selective Mott Phase}

\author{Yuekun Niu$^{1}$}
\email[]{ykniu@imu.edu.cn}%

\author{Yu Ni$^{2}$}

\author{Jia-Ming Wang$^{3}$}

\author{Zhong-Yi Lu$^{4}$}
\email[]{zlu@ruc.edu.cn}%

\author{Yun Song$^{5}$}
\email[]{yunsong@bnu.edu.cn}%

\author{Shiping Feng$^{5}$}
\email[]{spfeng@bnu.edu.cn}%

\affiliation{$^{1}$School of Physical Science and Technology $\&$ Inner Mongolia Key
Laboratory of Microscale Physics and Atomic Manufacturing, Inner Mongolia University,
Hohhot 010021,  China}

\affiliation{$^{2}$College of Physics and Electronic Information, Yunnan Normal University,
Kunming 650500, China}

\affiliation{$^{3}$Center for Materials Theory, Department of Physics and Astronomy, Rutgers
University, New Jersey 08854, USA}

\affiliation{$^{4}$Department of Physics, Renmin University of China, Beijing 100872, China}

\affiliation{$^{5}$School of Physics and Astronomy, Beijing Normal University, Beijing 100875,
China and \\
Department of Physics, Beijing Normal University, Zhuhai 519087, China}


\begin{abstract}
The orbital‑selective Mott phase (OSMP) is commonly described as a coexistence of localized and itinerant electrons within effectively decoupled orbitals, but emerging evidence for quasiparticle breakdown points to physics beyond this picture, whose microscopic origin remains unknown.
Using dynamical mean-field theory for the two-band Hubbard model, we show that the spin-flip and Ising-type components of Hund's coupling generate local spin excitations (LSEs). 
These LSEs couple electrons between different orbitals, renormalize quasiparticle lifetimes and binding energies, and thereby destroy well-defined quasiparticles in the OSMP. 
Removing these two components of Hund's coupling restores coherent quasiparticle behavior and fully decouples the charge dynamics of the two bands. 
Our results therefore identify electronic coupling to LSEs as the fundamental mechanism driving quasiparticle breakdown within the OSMP.
\end{abstract}

\pacs{71.30.+h, 71.27.+a, 71.10.-t}

\maketitle

\textit{Introduction}.
The orbital-selective Mott phase (OSMP), in which electrons in some orbitals localize while
those in others remain itinerant, is a fundamental manifestation of multiorbital
correlations that go beyond the single-band paradigm of Mott transitions
\cite{0Medici2009,7Georges2013,4Anisimov2002}. 
This exotic phase has attracted considerable interest because of its implications for strongly correlated electron systems
\cite{34Y2018,14Hallberg2020,15Song2020,17Wu2021,18Herbrych2021,Stepanov2024,2Kim2024,3Kugler2022,5Bouadim2009}.
In particular, it provides a natural framework for describing the coexistence of localized and itinerant electrons, which plays a central role in magnetism and charge transport \cite{25Koga2004,26Liebsch2003,27Shimoyamada2009}.
Traditionally, the OSMP is interpreted as two effectively decoupled subband systems,
namely, a gapped narrow band (NB) coexisting with a gapless wide band (WB) for a two-band model.
However, this straightforward picture has been challenged by angle-resolved photoemission spectroscopy (ARPES) measurements, which have revealed suppressed quasiparticle coherence in materials undergoing a metal-insulator transition (MIT) \cite{6Neupane2009,Vecchio2016,9Kim2018}, such as Ca$_{1.8}$Sr$_{0.2}$RuO$_{4}$, and correlation-driven Fermi surface (FS) reconstruction in FeTe$_{1-x}$Se$_x$ \cite{Huang2022}.
Taken together, these findings suggest the presence of unexpected interorbital coupling
\cite{6Neupane2009,Vecchio2016,9Kim2018,Huang2022}.
They therefore raise an unresolved question: What microscopic mechanism couples the charge dynamics of the seemingly decoupled localized and itinerant electrons?

Hund's coupling is central to this question, yet the microscopic mechanism by which it couples the charge dynamics of the two bands remains elusive.
After decades of intensive investigation, the role of Hund's coupling in stabilizing the OSMP and tuning the Mott transition has been well established \cite{41Haule2009,42Werner2012,43Kune2012,39Yin2011,40Luca2011,62Huang2012,64Stadler2015}.
In particular, the interplay among Hund's coupling, Coulomb repulsion, and orbital bandwidth in governing the transition has also been extensively investigated
\cite{44Biermann2005,46Luca2011,47Liebsch2004,48Ferrero2005}. 
Moreover, pioneering studies within the framework of dynamical mean-field theory (DMFT)
\cite{36Georges1996,37Kotliar2006,38Vollhardt2012} have shown that interorbital spin fluctuations in the paramagnetic OSMP can induce non-Fermi-liquid behavior \cite{44Biermann2005,Yin2012}.
However, these conventional analyses typically rely on qualitative descriptors, such as the Mott gap and quasiparticle coherence, to characterize the OSMP
\cite{28Zhang2012,29Dagotto2011,30Luca2005,31Strigari2013,32Yu2013,33Lee2010}; 
consequently, a quantitative measure of the degree of interorbital coupling remains lacking.
In recent work, we introduced the local two-qubit fidelity (LTQF), a quantitative metric that clearly distinguishes the metallic state ($L_{\rm o}=0$) from the Mott insulating state ($L_{\rm o}=0.5$), thereby providing a quantitative characterization of the Mott MIT in the OSMP \cite{345Niu2023,35Niu2024}. 
For a conventional decoupled OSMP, the LTQF is restricted to these integer and half-integer values \cite{345Niu2023,35Niu2024}. 
By contrast, when the full Hund's coupling is retained, we identified non-half-integer LTQF values within the OSMP and demonstrated that the NB exhibits a quantum-entangled state, providing direct quantitative evidence for interorbital coupling in the charge degrees of freedom \cite{35Niu2024}.
Nevertheless, the physical origin of this anomaly and whether it reflects novel low-energy excitations \cite{R2017,Liu2018,56Aucar2024,60Wang2025} remain unclear.
These open questions are further underscored by recent studies of dynamical effects arising from local interorbital correlations, including in-gap Hund excitations \cite{54Aucar2021,55Sroda2023} and non-Fermi-liquid transport in Hund metals
\cite{53Kugler2019}, all of which point to a common picture of interorbital spectral reconstruction.

In this Letter, we show that local spin excitations (LSEs), generated by the
spin-flip ($J_{\rm sf}$) and Ising-type ($J_{\rm z}$) components of Hund's coupling, play a
crucial role in the physics of the OSMP. 
These LSEs mediate dynamical interorbital charge transfer, thereby renormalizing the quasiparticle spectrum and destroying coherent quasiparticles.
This breakdown is quantitatively reflected in the emergence of non-half-integer LTQF values.
In contrast, removing the spin-flip and Ising-type Hund's coupling components eliminates the LSEs, restores coherent quasiparticle behavior, and returns the system to the conventional decoupled OSMP characterized by integer and half-integer LTQF values.

\textit{Models and methods.\label{sec:mm}}
Our calculations are based on the half‑filled, nonhybridized two‑band Hubbard model defined on a Bethe lattice \cite{73Roth1966,74Ole1983},
\begin{eqnarray}\label{hubbard}
H=-\sum_{\langle i j \rangle\alpha\sigma}t_{\alpha}d_{i\alpha\sigma}^\dag d_{j\alpha\sigma}
-\mu\sum_{i \alpha\sigma}d_{i \alpha\sigma}^\dag d_{i \alpha\sigma}
+ H_{\rm int}[d_{\alpha\sigma}],~~~~
\end{eqnarray}
with the interaction term $H_{\rm int}[d_{\alpha\sigma}]$ given by
\begin{eqnarray}\label{hubbard_H_int}
&H&_{\rm int}[d_{\alpha\sigma}]= U\sum_{i\alpha}n_{i\alpha\uparrow}n_{i\alpha\downarrow}
+U^{\prime}\sum_{i,\alpha\neq\alpha^{\prime}}n_{i\alpha\uparrow}
n_{i\alpha^{\prime}\downarrow}	\nonumber\\
&+&(U^{\prime} - J_{\rm z})\sum_{i\alpha<\alpha^{\prime}\sigma}n_{i\alpha\sigma}
n_{i\alpha^\prime\sigma}+J_{\rm ph}\sum_{i,\alpha\neq \alpha^{\prime}}
d_{i \alpha\uparrow}^\dag
d_{i \alpha\downarrow}^{\dag}d_{i \alpha^{\prime}\downarrow}d_{i \alpha^{\prime}\uparrow}
\nonumber\\
&+&J_{\rm sf}\sum_{i,\alpha\neq \alpha^{\prime}}d_{i \alpha\uparrow}^\dag
d_{i \alpha^{\prime}\downarrow}^{\dag}d_{i \alpha\downarrow}d_{i \alpha^{\prime}\uparrow},
\end{eqnarray}
where $d_{i\alpha\sigma}^\dag(d_{i\alpha\sigma})$ is the creation (annihilation) operator
for an electron in the orbital $\alpha=1,~2$ with spin $\sigma$ at site $i$,
$n_{i\alpha\sigma}$ is the occupation number operator of electrons in the orbital $\alpha$
at site $i$, $t_{\alpha}$ denotes the electron hopping amplitude in the orbital $\alpha$,
$U$ ($U^{\prime}$) represents the intraorbital (interorbital) Coulomb interaction, Hund's coupling terms consist of the Ising-type term $J_{\rm z}$, the spin-flip term
$J_{\rm sf}$, and the pair-hopping term $J_{\rm ph}$, and $\mu$ is the chemical potential.
The summation $\langle ij \rangle$ is taken over all sites $i$ and, for each site $i$,
is restricted to its nearest-neighbor (NN) sites $j$. Throughout this paper, we fix
$J_{\rm z}$, $J_{\rm sf}$, and $J_{\rm ph}$ such that $J_{\rm z}=J_{\rm sf}=J_{\rm ph}=J$,
while the relationship $U=U^{\prime}+2J$ is chosen to maintain spin rotational symmetry \cite{75Castellani1978,76Raymond1997}. In particular, the bandwidth ratio is defined as $R=t_{2}/t_{1}$, where $t_{1}$ and $t_{2}$ are the electron hopping
amplitudes in the WB and NB, respectively. 
Moreover, the hopping amplitude $t_{1}$ in the WB is set to the energy unit.

The above two-band model (\ref{hubbard}) can be mapped onto an effective single-impurity
model,
\begin{eqnarray}\label{anderson}
H_{\rm imp}=\hat{H}_{0}+\hat{H}_{\rm int}[d_{\alpha\sigma}],
\end{eqnarray}
with the noninteracting term,
\begin{eqnarray}\label{anderson_H0}
H_{0}&=&\sum_{m \alpha\sigma}\varepsilon_{m \alpha}c_{m \alpha\sigma}^\dag
c_{m \alpha\sigma}-\sum_{\alpha\sigma}\mu d_{\alpha\sigma}^\dag d_{\alpha\sigma}
\nonumber\\
&+&\sum_{m \alpha\sigma}V_{m \alpha}(c_{m \alpha\sigma}^\dag d_{\alpha\sigma}
+d_{\alpha\sigma}^\dag c_{m \alpha\sigma}) ,
\end{eqnarray}
where $d_{\alpha\sigma}^{\dag}$ ($d_{\alpha\sigma}$) is the creation (annihilation) operator at the impurity site for orbital $\alpha$ with spin $\sigma$, and $c_{m\alpha\sigma}^{\dag}$ ($c_{m\alpha\sigma}$) is the corresponding operator in the conduction band. 
The {\it impurity site} and {\it conduction band} are coupled
to each other via the effective parameters $\varepsilon_{m\alpha}$ and $V_{m\alpha}$,
which are determined by the self-consistent DMFT calculations. The above mapping becomes
exact for an infinite lattice \cite{77Georges1992}.

The local Green's function $\mathcal{G}_{\alpha\sigma}(i\omega_{n})$ of the effective single-impurity model (\ref{anderson}) on a Bethe lattice with a semicircular density of states (DOS) is obtained from a single-site impurity problem supplemented by the self-consistent relation \cite{78caffarel1994,79Laloux1994} (see Supplemental Material \cite{sm}). 
The self-consistent relation ensures that the on-site (local) component of the Green's function, $\mathcal{G}_{ii}(i\omega_{n})=\sum_{k}\mathcal{G}(k,i\omega_{n})$, coincides with the Green's function $\mathcal{G}(i\omega_{n})$ derived from the effective action.
Solving this quantum impurity problem is the central numerical step in DMFT, and a variety of powerful methods have been developed for this task
\cite{49Bulla2008,50Gull2011}. 
In the present work, we employ the Lanczos exact diagonalization technique as the impurity solver, since it avoids the fermion sign problem and can flexibly adapt to Hamiltonians with multiple orbitals, low symmetry, and spin-orbit coupling.
It yields real-frequency spectra and serves as an efficient, high-precision exact diagonalization solver. 
We perform calculations at zero temperature, where $\beta$ is treated as a large but finite parameter to discretize the imaginary-time interval, thereby approaching the zero-temperature limit \cite{78caffarel1994}.
We use the bath size $n_{\mathrm{b}}=3$ and $\beta=512$.

\begin{figure}
\centering
\includegraphics[width=0.47\textwidth]{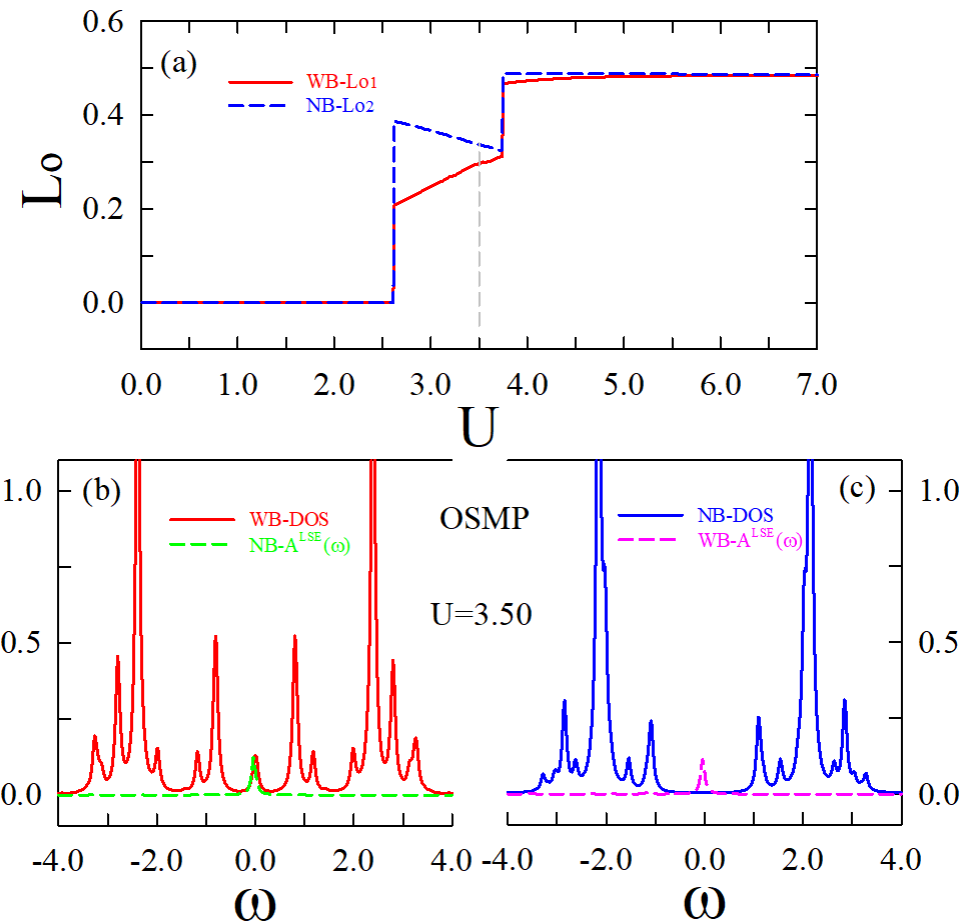}
\caption{\label{fig:ltqf_dos_eh} LSEs in OSMP: (a) The LTQF of the WB (red solid line)
and NB (blue dotted line) are plotted against Coulomb interaction $U$. The Mott critical
point of WB is $U_{c1}=3.74$, while that of NB is $U_{c2}=2.62$; (b) DOS (red solid line)
of WB and the LSE spectrum $A^{lse}(\omega)$ (green dotted line) of NB at $U=3.50$ (inside the OSMP); (c) The DOS (blue solid line) of NB and the LSE spectrum (pink dotted line) of WB at
the same $U$. The model parameters are $J=U/4$ and $R=0.5$. The LSE peaks at the Fermi level in the NB Mott gap provide a low-energy channel for interorbital charge fluctuations.}
\end{figure}

\textit{Emergence of LSEs and Interorbital Coupling}.
Recent work has established the LTQF as a quantitative characterization of the Mott transition in multiorbital systems \cite{345Niu2023,35Niu2024}. 
In particular, the expected LTQF values are $0$ (fully metallic band) or $0.5$ (perfectly
localized band) for a decoupled OSMP. 
However, when $J\neq 0$, a notable deviation from this quantization criterion occurs \cite{35Niu2024}, yielding non-half-integer LTQF values that signal correlation between the WB and NB.
To substantiate this point, Fig. \ref{fig:ltqf_dos_eh}a displays the orbital-resolved LTQF as a function of $U$ for $R=0.5$ and $J=U/4$.
Within the OSMP ($U_{\rm c2}=2.62<U<U_{\rm c1}=3.74$),
the LTQF values deviate significantly from their quantized ideal values \cite{35Niu2024}, directly signaling interorbital coupling.
We now elucidate the microscopic origin of this interorbital correlation by linking it to LSEs. 
The Green's functions $G_{1}(\omega)$ and $G_{2}(\omega)$ for the WB and NB are analytically derived via the equation of motion method (see Supplemental Material \cite{sm}).
The resulting expressions reveal that $G_{\alpha}(\omega)$ explicitly depends on Hund's coupling correlation functions of the other orbital, demonstrating that LSEs act as a dynamical link between the two bands.
To better characterize these exotic quasiparticle excitations, we introduce the LSE spectral function defined as
\begin{equation}\label{eph}
A_{\alpha}^{\rm LSE}(\omega)=-\frac{1}{\pi}{\rm Im}\left \langle
b_{\alpha}^{\dag}{\frac{1}{\omega-H+i\delta^{+}}b_{\alpha}}\right\rangle,
\end{equation}
where $\delta$ is an infinitesimal and the composite operator
$b_{\alpha}^{\dag}=d_{\alpha\downarrow}^{\dag}d_{\alpha\uparrow}$. 
It should be emphasized that this operator $b_{\alpha}^{\dag}$ explicitly describes a local spin-flip excitation, which simultaneously creates an electron-hole pair in orbital $\alpha$ across the spin channel. 
This LSE intrinsically couples the spin and charge degrees of freedom of the electron, and its features are different from those of pure magnetic or charge excitations.

\begin{figure}
\centering
\includegraphics[width=0.47\textwidth]{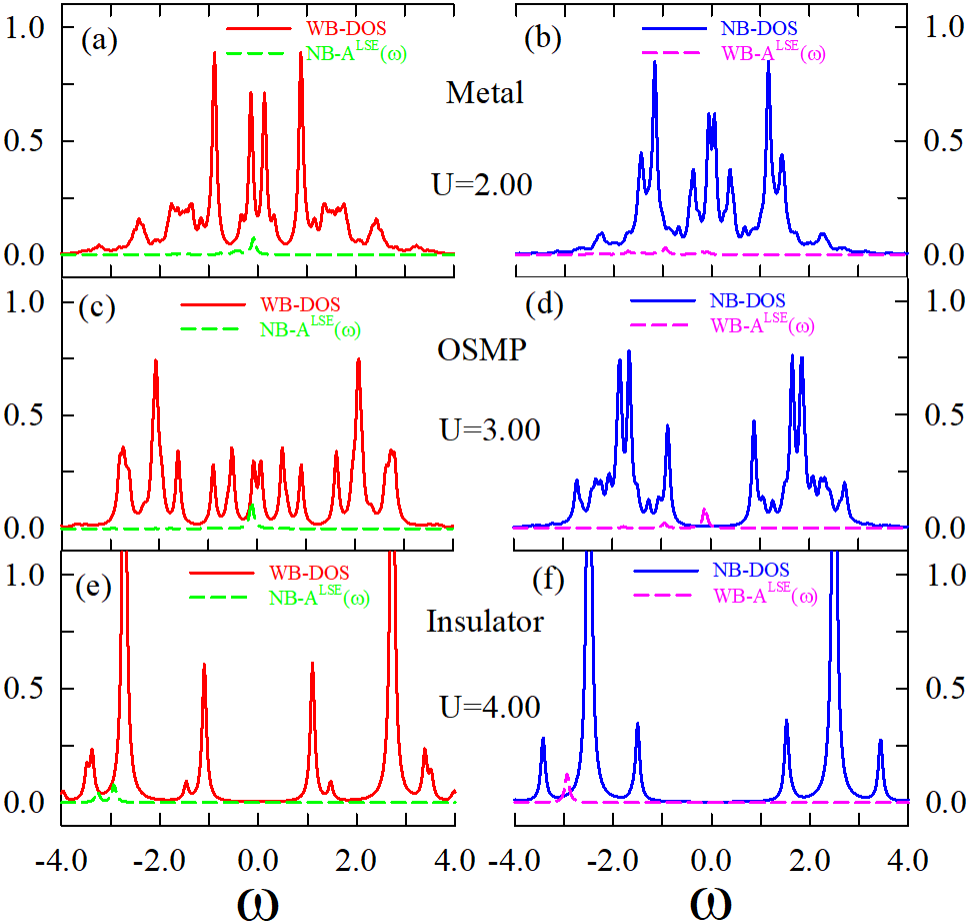}
\caption{\label{fig:dos_eh} Evolution of the LSE spectra across the Mott transition. Left column: the LSE spectrum of NB (green dashed lines) superimposed on the DOS
of WB (red solid lines); Right column: the LSE spectrum of WB (pink dashed lines)
superimposed on DOS of NB (blue solid lines). (a) and (b) Metallic phase at $U=2.00$;
(c) and (d) OSMP at $U=3.00$; (e) and (f) Insulating phase at $U=4.00$. The model
parameters are $J=U/4$ and $R=0.5$. The LSE spectral weight shifts to the Fermi level
only in the OSMP, signaling the emergence of the interorbital coupling. }
\end{figure}

To reveal how these LSEs manifest themselves in the interacting system, we compute $A_{\alpha}^{\rm LSE}(\omega)$ within DMFT and display it alongside the DOS in Figs. \ref{fig:ltqf_dos_eh}b and \ref{fig:ltqf_dos_eh}c for $U=3.50$, deep within the OSMP.
The DOS is calculated as $\rho_{\alpha}(\omega)=-\frac{1}{\pi}{\rm Im}G_{\alpha}(\omega+i\eta)$, where $\eta$ is an energy broadening factor.
It is demonstrated that both the localized NB and the itinerant WB exhibit sharp LSE resonances precisely at the Fermi level (FL) $\omega=0$, which constitutes the central finding. Crucially, the LSE resonance of WB resides inside the Mott gap of NB
[Fig. \ref{fig:ltqf_dos_eh}c], while the LSE of NB coexists with the coherent quasiparticle
peak of WB [Fig. \ref{fig:ltqf_dos_eh}b]. These LSEs at the FL provide a low-energy channel
for dynamic interorbital charge fluctuations. The originally insulating NB acquires a small amount of itinerant spectral weight, while the WB loses part of its coherent quasiparticle weight. 
This self-consistent picture demonstrates how LSEs act as dynamical intermediaries, transferring spectral weight between bands and providing a natural microscopic explanation for the observed non-half-integer LTQF values.

To confirm this mechanism, Fig. {\ref{fig:dos_eh}} presents the evolution of the LSE spectrum
across the phase transition. 
If the LSE-induced spectral weight transfer is indeed responsible for the non-half-integer LTQF values, the LSE spectral weight should be concentrated at low energies only within the OSMP and absent in both the metallic and insulating limits.
In the metallic phase ($U=2.0$), the LSE spectra for both bands
are weak and lie below the FL [see Figs. \ref{fig:dos_eh}a and \ref{fig:dos_eh}b].
Upon entering the OSMP ($U=3.0$), the LSE spectral weight shifts dramatically
to the FL and condenses into sharp resonances for both bands, coincident with the onset of the LTQF anomaly [Figs. \ref{fig:dos_eh}c and \ref{fig:dos_eh}d].
This indicates that the emergence of LSEs is phase-selective, appearing precisely within the OSMP, where the non-half-integer LTQF values signal interorbital coupling.
In the fully insulating phase ($U=4.0$), the LSEs at FL vanish completely [Figs. \ref{fig:dos_eh}e and \ref{fig:dos_eh}f], consistent with the frozen charge dynamics of a Mott insulator. This clear correlation between the emergence of LSEs at the FL and the LTQF anomaly firmly establishes
LSEs as the key agents of interorbital coupling.
This evolution also indicates that both the concentration and the energy position of LSEs are
intimately linked to the Mott MIT.

\begin{figure}
\centering
\includegraphics[width=0.47\textwidth]{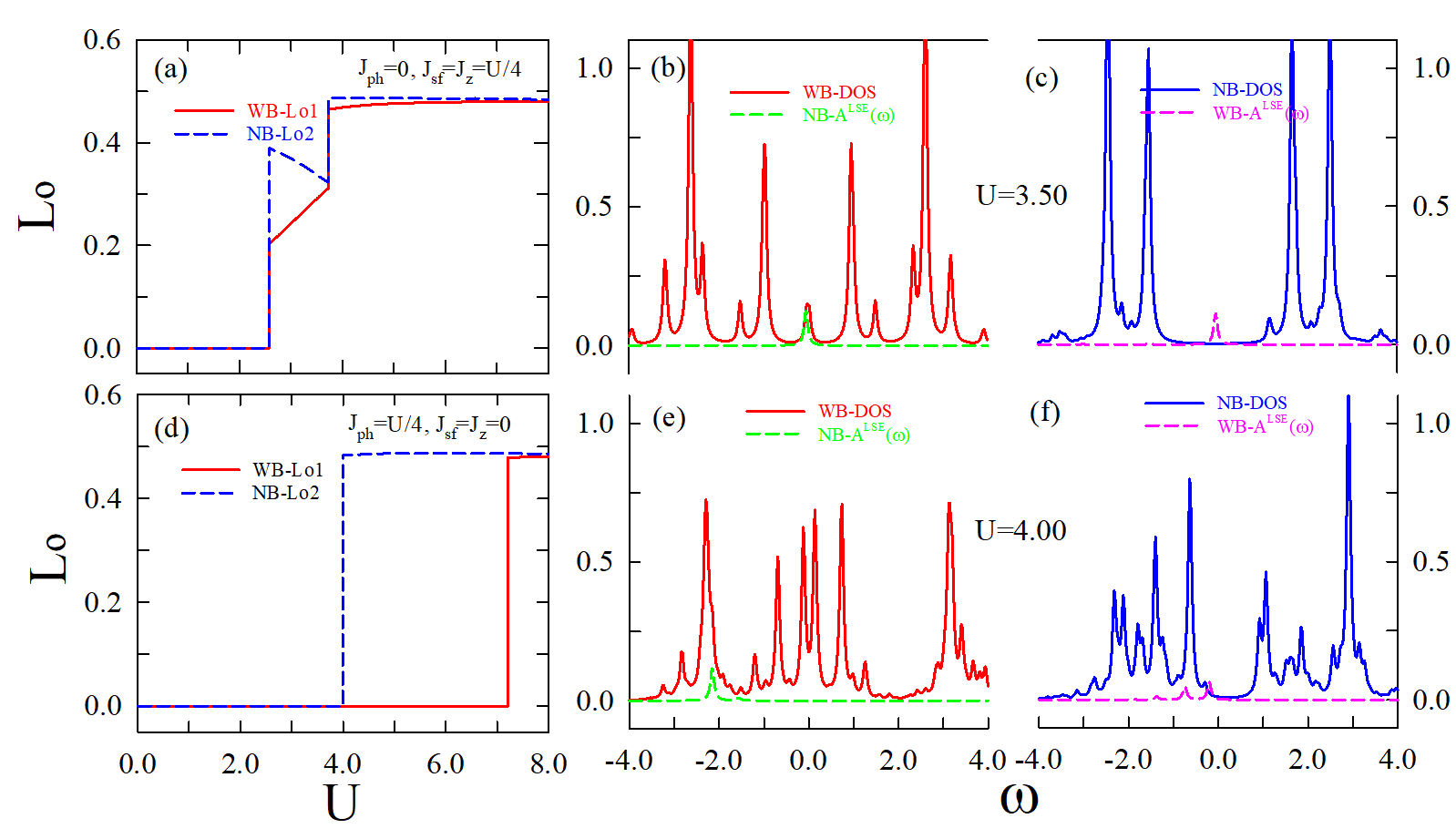}
\caption{\label{fig:LoR} Role of Hund's coupling components in generating LSEs. (a)-(c) Results for $J_{\rm ph}=0$ and $J_{\rm sf}=J_{\rm z}=U/4$. (a) LTQF versus $U$
showing non-half-integer values in the OSMP ($U_{\rm c1}=3.74, U_{\rm c2}=2.58$); (b) LSE
spectrum of the NB (green dashed) and DOS of the WB (red solid); (c) LSE spectrum of the WB
(pink dashed) and DOS of the NB (blue solid); (d)-(f) Results for $J_{\rm ph}=U/4$ and
$J_{\rm sf}=J_{\rm z}=0$. (d) LTQF versus $U$ showing exact $L_{\rm o1}=0$ and $L_{\rm o2}=0.5$
in the OSMP ($U_{\rm c1}=7.21, U_{\rm c2}=4.00$); (e) and (f) Complete absence of LSE spectra,
confirming the conventional decoupled OSMP. All results are obtained for $R=0.5$. The
comparison demonstrates that $J_{\rm sf}$ and $J_{\rm z}$ are essential for generating LSEs and interorbital coupling.}
\end{figure}

\textit{Key Role of $J_{\rm sf}$ and $J_{\rm z}$}.
To unambiguously verify this causal scenario, we now isolate the Hund’s coupling components.
We performed two complementary numerical calculations: (i) With $J_{\rm ph}=0$ and $J_{\rm sf}=J_{\rm z}=U/4$, the OSMP ($U_{c1}=3.74$,$U_{c2}=2.58$)
displays robust non-half-integer LTQF values [see Fig. \ref{fig:LoR}a] and clear LSEs at the FL
in both bands [Figs. \ref{fig:LoR}b and \ref{fig:LoR}c]. This indicates that the spin-flip
and Ising-type components alone are sufficient to generate interorbital coupling. (ii) With $J_{\rm ph}=U/4$ and $J_{\rm sf}=J_{\rm z}=0$, the LSE spectra
vanish entirely [Figs. \ref{fig:LoR}e and \ref{fig:LoR}f], and the LTQF values return to their
ideal values of $L_{o1}=0$ and $L_{o2}=0.5$ throughout the entire OSMP ($U_{c1}=7.21$, $U_{c2}=4.00$) [Fig. \ref {fig:LoR}d]. This comparison demonstrates that the spin-flip and Ising-type terms,
and the LSEs they generate, are essential for the breakdown of the conventional OSMP picture.
With these terms absent, the WB retains its fully coherent quasiparticle character, whereas the NB forms a purely localized Mott insulator. 
The notable shift in the critical $U_{\rm c}$ values between the two cases further confirms that $J_{\rm sf}$ and $J_{\rm z}$ are much more effective than $J_{\rm ph}$ alone in generating strong correlations and coupling the orbitals.
The OSMP is thus not a mere coexistence of distinct phases, but an emergent many-body state
whose intrinsic properties are critically governed by these specific components of Hund's coupling.

\textit{Conclusions and discussion}.
Our study establishes that LSEs, induced by the spin-flip and Ising-type components of Hund's coupling, are the fundamental microscopic mechanism mediating quasiparticle breakdown in the OSMP.
These LSEs act as dynamical intermediaries, renormalizing the charge self-energy and
redistributing spectral weight between the WB and NB. 
This process effectively eliminates the sharp distinction between coherent itinerant quasiparticles and fully localized Mott excitations. 
Instead, it yields a dynamically coupled state in which the coherent quasiparticle weight of the WB is partially transferred to the NB, such that the WB becomes less coherent while the nominally insulating NB gains incoherent spectral weight with itinerant character inside its Mott gap.
The non-half-integer LTQF values are a direct and quantitative signature of this LSE-mediated
quasiparticle breakdown.

Our results provide a robust microscopic framework for understanding several key anomalous phenomena in multiorbital correlated systems. In particular, the suppressed quasiparticle coherence and pervasive non-Fermi-liquid transport features in Hund metals \cite{44Biermann2005} can be largely ascribed to the LSE-mediated spectral weight redistribution uncovered in this work.
While previous studies have well documented Hund’s coupling as a key source of orbital differentiation \cite{7Georges2013}, our work further pinpoints its spin-flip and Ising-type components constitute the fundamental microscopic origin of these interorbital correlations.
This refined classification clarifies the difference between pure Ising-type and full Hund's coupling proposed in prior literature \cite {Song2017} and is consistent with recent theories addressing complex orbital entanglement within the OSMP \cite{60Wang2025}.
Furthermore, the correlation-driven Fermi surface reconstruction observed via ARPES on FeTe$_{1-x}$Se$_x$ \cite{Huang2022} can be attributed to this same interorbital coupling channel, which offers a new paradigm for interpreting its microscopic origin.
Most crucially, this LSE-driven unified microscopic picture simultaneously accounts for the ARPES results of metallic V$_2$O$_3$ reported in Ref. \cite{Vecchio2016}, where both $a_{\rm 1g}$ and $e_{\rm g}^{\rm \pi}$ orbitals are experimentally confirmed to contribute to the FS.

In summary, our work demonstrates that the OSMP is not merely a coexistence of localized and itinerant electrons. 
Rather, the spin-flip and Ising-type components of Hund's coupling generate local spin excitations that dynamically intertwine the orbitals, challenging the conventional orbital-resolved quasiparticle picture. 
This suggests viewing the OSMP as an emergent correlated phase in which Hund's coupling intrinsically couples the spin and charge degrees of freedom across orbitals.

\textit{Acknowledgements}.
This work has been supported by the Scientific Research Foundation for Youth Academic Talent of Inner Mongolia University under Grant No. 10000-23112101/010 and the Natural Science Foundation of Inner Mongolia Autonomous Region under Grant No. 21200-52531050. S.F. acknowledges the
support from the National Key Research and Development Program of China under Grant
Nos. 2023YFA1406500 and 2021YFA1401803 and the National Natural Science Foundation of China
under Grant No. 12274036.

\end{document}


\title{Supplemental Material:Local Spin Excitations Mediate Quasiparticle Breakdown in the Orbital-Selective Mott Phase}

\author{Yuekun Niu$^{1}$}
\email[]{ykniu@imu.edu.cn}%

\author{Yu Ni$^{2}$}

\author{Jia-Ming Wang$^{3}$}

\author{Zhong-Yi Lu$^{4}$}
\email[]{zlu@ruc.edu.cn}%

\author{Yun Song$^{5}$}
\email[]{yunsong@bnu.edu.cn}%

\author{Shiping Feng$^{5}$}
\email[]{spfeng@bnu.edu.cn}%

\affiliation{$^{1}$School of Physical Science and Technology, $\&$ Inner Mongolia Key
Laboratory of Microscale Physics and Atomic Manufacturing, Inner Mongolia University,
Hohhot 010021,  China }

\affiliation{$^{2}$College of Physics and Electronic Information, Yunnan Normal University,
Kunming 650500, China}

\affiliation{$^{3}$Center for Materials Theory, Department of Physics and Astronomy, Rutgers
University, New Jersey 08854, USA}

\affiliation{$^{4}$Department of Physics, Renmin University of China, Beijing 100872, China}

\affiliation{$^{5}$School of Physics and Astronomy, Beijing Normal University, Beijing 100875,
China and\\
Department of Physics, Beijing Normal University, Zhuhai 519087, China}

\maketitle

These supplemental materials consist of the details of analytic calculations as well as additional numerical results supporting the findings presented in the main text.

\section{DMFT self-consistent relationship}\label{app:dmft}
The local Green's function on a Bethe lattice with a semicircular DOS is obtained from
a single-site impurity problem supplemented by the self-consistent relation
\cite{78caffarel1994,79Laloux1994} as,
\begin{equation}\label{self-consistency}
\mathcal{G}_{0\alpha\sigma}(i\omega_{n})={\frac{1}{ i\omega_{n}+\mu-t_{\alpha}^{2}
\mathcal{G}_{\alpha\sigma}(i\omega_{n})}},
\end{equation}
where $\mathcal{G}_{0}$ is the bare Green's function. The self-consistent relation
ensures that the on-site (local) component of the Green's function $[\mathcal{G}_{ii}(i\omega_{n})=\sum_{k}\mathcal{G}(k,i\omega_{n})]$ coincides with
the Green's function $\mathcal{G}(i\omega_{n})$ calculated from the effective action.

To quantitatively identify the Mott transition, we employ LTQF
\cite{345Niu2023,35Niu2024},
\begin{equation}\label{Io}
L_{{\rm o}\alpha}=-\frac{1}{\beta}\sum_{n=-\infty}^{\infty}e^{i\omega_{n}0^{+}}
\mathcal{G}_{\alpha}(i\omega_{n})\langle \Phi_{imp}^{o\alpha}|\hat{P}|
\Phi_{imp}^{o\alpha}\rangle,
\end{equation}
with $|\Phi_{imp}^{o\alpha}\rangle$ given by,
\begin{equation}
|\Phi_{imp}^{o\alpha}\rangle=p_{\alpha1}|0\rangle+p_{\alpha2}|\uparrow\rangle
+p_{\alpha3}|\downarrow\rangle +p_{\alpha4}|\uparrow\downarrow\rangle,\nonumber
\end{equation}
where $\hat{P}$ is the net spin projection operator for the impurity site with $\langle0|\hat{P}|0\rangle=\langle\uparrow\downarrow|\hat{P}|\uparrow\downarrow\rangle=0$, $\langle\uparrow|\hat{P}|\uparrow\rangle=1$, and
$\langle\downarrow|\hat{P}|\downarrow\rangle=-1$.
$|\Phi_{imp}^{o\alpha}\rangle$ represents the ground-state wave function of a single
impurity with an interaction strength of $U$. The ground-state is represented
by the superposition of spin-up, spin-down, zero, and double-occupied states. For a
decoupled OSMP, the expected LTQF values are $0$ (fully metallic band) or $0.5$ (perfectly
localized band). Our calculations for $J\neq0$ show significant deviations from these ideals,
yielding non-half-integer LTQF values that exhibit a certain correlation between WB and NB
\cite{35Niu2024}.

\begin{figure}
\centering
\includegraphics[width=0.5\textwidth]{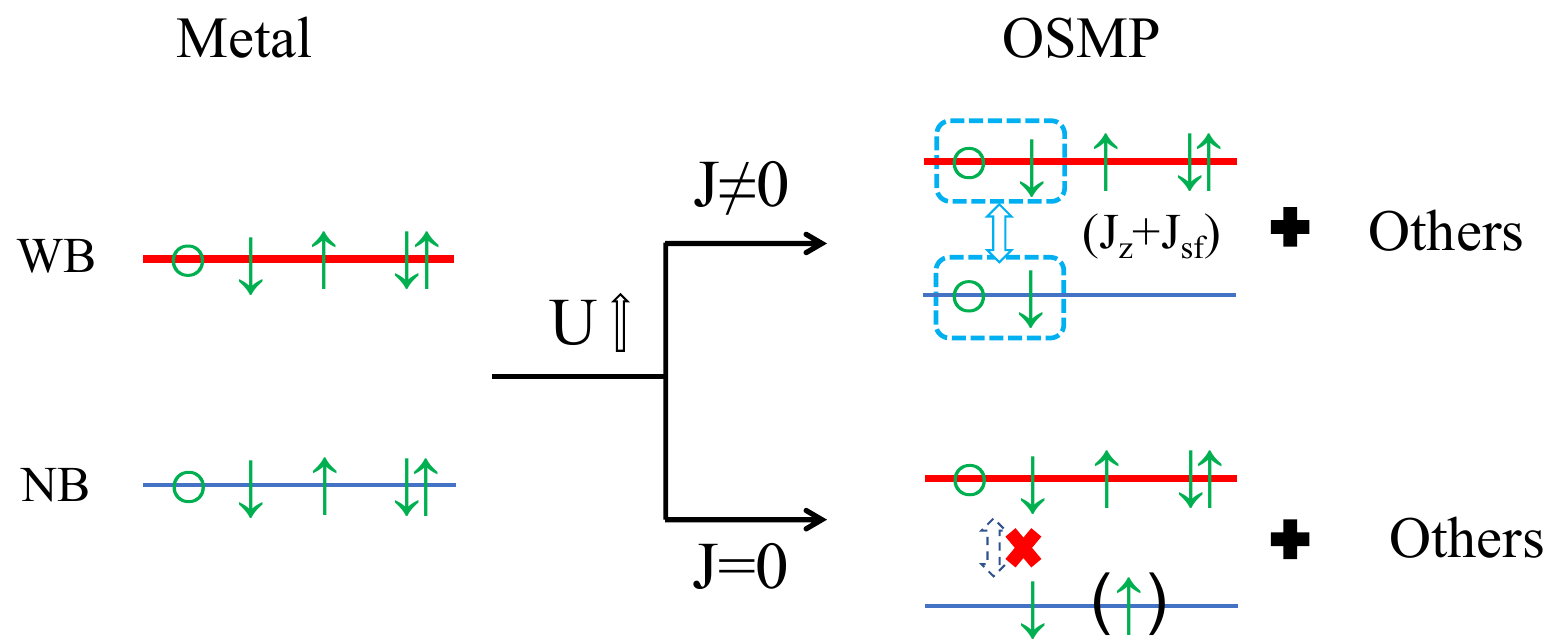}
\caption{\label{fig:0} Schematic diagram of LSE bound states. WB is represented by a thicker
red line and NB by a thinner blue line. The dashed box encloses the LSE bound state.
The “others” in the figure correspond to other potential low-energy excited states.}
\end{figure}

\section{Green's function equations of motion}\label{app:gfem}
To gain analytical insight into the origin of the non-half-integer LTQF values, we derive
the equations of motion for the impurity Green's function.
\begin{eqnarray}\label{hf}
<<[d_{\alpha\sigma},H_{int}]|d_{\alpha\sigma}^{\dag}>>=&U
<<d_{\alpha\sigma}n_{\alpha\bar{\sigma}}|d_{\alpha\sigma}^{\dag}>> + U^{\prime} \sum\limits_{\alpha\neq\alpha^{\prime}}
<<d_{\alpha\uparrow}n_{\alpha^{\prime}\downarrow}|d_{\alpha,\sigma}^{\dag}>>+ J_{sf}\sum\limits_{\alpha\neq\alpha^{\prime}}
<<d_{\alpha^{\prime}\downarrow}^{\dag}d_{\alpha\downarrow}d_{\alpha^{\prime}\uparrow}|
d_{\alpha\sigma}^{\dag}>> \nonumber  \\
+& J_{ph}\sum\limits_{\alpha\neq\alpha^{\prime}}<<(d_{\alpha\downarrow}^{\dag}d_{\alpha^{\prime}\downarrow}d_{\alpha^{\prime}\uparrow}
+d_{\alpha\uparrow}^{\dag}d_{\alpha^{\prime}\downarrow}d_{\alpha^{\prime}\uparrow})|
d_{\alpha\sigma}^{\dag}>> + (U^{\prime}-J_{z})\sum\limits_{\alpha<\alpha^{\prime}}<<d_{\alpha\sigma}
n_{\alpha^{\prime}\sigma}|d_{\alpha\sigma}^{\dag}>>.
\end{eqnarray}
With the help of the following Hartree-Fock approximation on the correlation terms,
\begin{eqnarray}
<<d_{\alpha\sigma}n_{\alpha\bar{\sigma}}|d_{\alpha\sigma}^{\dag}>>
&\approx& <n_{\alpha\bar{\sigma}}><<d_{\alpha\sigma}|d_{\alpha\sigma}^{\dag}>>,\\
<<d_{\alpha\uparrow}n_{\alpha^{\prime}\downarrow}|d_{\alpha,\sigma}^{\dag}>>
&\approx& <n_{\alpha^{\prime}\downarrow}><<d_{\alpha\uparrow}|d_{\alpha,\sigma}^{\dag}>>,\\
<<d_{\alpha\sigma}n_{\alpha^{\prime}\sigma}|d_{\alpha\sigma}^{\dag}>>
&\approx& <n_{\alpha^{\prime}\sigma}><<d_{\alpha\sigma}|d_{\alpha\sigma}^{\dag}>>,~~~~~
\end{eqnarray}
the equation (\ref{hf}) can be expressed approximately as,
\begin{eqnarray}\label{gfbs}
&[\omega_{m\alpha c}-U<n_{\alpha\bar{\sigma}}>-(U^{\prime}
-J_{z})<n_{\alpha^{\prime}\sigma}>]<<d_{\alpha\sigma}|d_{\alpha\sigma}^{\dag}>>-U^{\prime}<n_{\alpha^{\prime}\downarrow}><<d_{\alpha\uparrow}|d_{\alpha\sigma}^{\dag}>> \nonumber \\
&+J_{sf}<d_{\alpha^{\prime}\downarrow}^{\dag}d_{\alpha^{\prime}\uparrow}>
<<d_{\alpha\downarrow}|d_{\alpha\sigma}^{\dag}>>=1, ~~~~
\end{eqnarray}
where $<<d_{\alpha\sigma}|d_{\alpha\sigma}^{\dag}>>=G_{\alpha\sigma}(\omega)$, and
then the Green's functions $G_{1}(\omega)$ for WB (the orbital $1$) and $G_{2}(\omega)$
for NB (the orbital $2$) are derived as,
\begin{eqnarray}
G_{1}(\omega)&=&G_{1\downarrow}(\omega)+G_{1\uparrow}(\omega) \nonumber \\
&=&\frac{1}{\omega_{m1c}-U\langle n_{1\uparrow}\rangle
-(U^{\prime}-J_{z})\langle n_{2\downarrow}\rangle
+J_{sf}\langle d_{2\downarrow}^{\dag}d_{2\uparrow}\rangle} \nonumber \\
&+&\frac{1}{\omega_{m1c}-U\langle n_{1\downarrow}\rangle
-(U^{\prime}-J_{z})\langle n_{2\uparrow}\rangle-U^{\prime}\langle n_{2\downarrow}\rangle},\label{wbgf}\\
G_{2}(\omega)&=&G_{2\downarrow}(\omega)+G_{2\uparrow}(\omega) \nonumber \\
&=&\frac{1}{\omega_{m2c}-U\langle n_{2\uparrow}\rangle-(U^{\prime}
-J_{z})\langle n_{1\downarrow}\rangle+J_{sf}\langle d_{1\downarrow}^{\dag}
d_{1\uparrow}\rangle} \nonumber \\
&+&\frac{1}{\omega_{m2c}-U\langle n_{2\downarrow}\rangle
-(U^{\prime}-J_{z})\langle n_{1\uparrow}\rangle
-U^{\prime}\langle n_{1\downarrow}\rangle},~~~\label{nbgf}
\end{eqnarray}
respectively,
\begin{figure}
\centering
\includegraphics[width=0.6\textwidth]{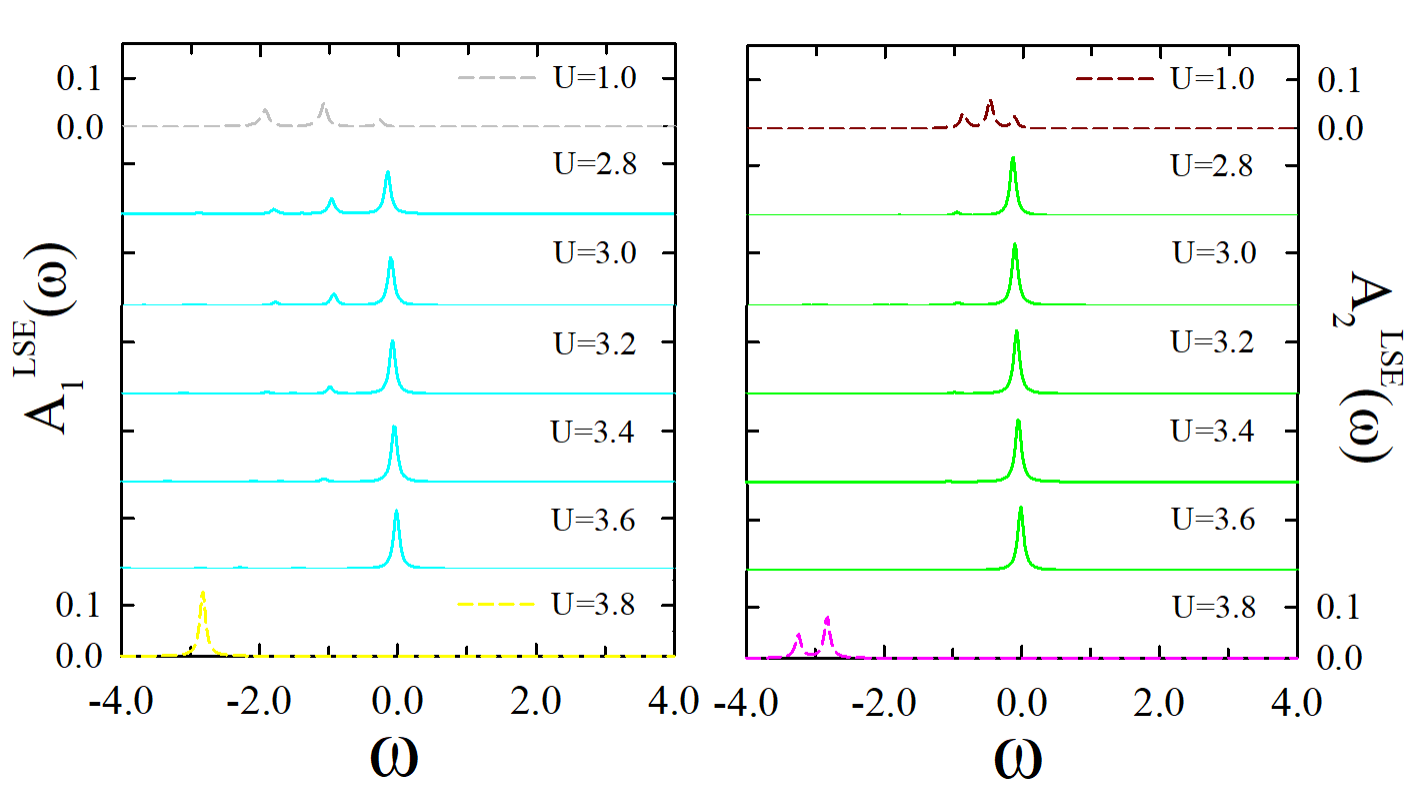}
\caption{\label{fig:AU} Effect of the Coulomb interaction U for the excitation spectrum
of LSE $A_{\alpha}^{\rm LSE}(\omega)$ in different phases. The metal phase for $U=1.0$
(gray and brown dashed lines), OSMP for $U=2.80, 3.00, 3.20, 3.40,3.60$ (cyan and green
solid lines), the insulating phase for $U=3.80$ (yellow and pink dashed lines), where the
spectrum has been normalized by summation. The model parameters are $J=U/4$ and $R=0.5$.}
\end{figure}
where $\omega_{m\alpha c}=\omega+\mu- {V_{m\alpha}^{2}}/{(\omega-\varepsilon_{m\alpha})}$.
Crucially, $G_{1}(\omega)$ depends not only on the NB occupation
$\langle n_{2}\rangle$ but also on the correlation
$\langle d_{2\downarrow}^{\dag}d_{2\uparrow}\rangle$, a correlation function whose
existence is generated by the spin-flip term $J_{\rm sf}$. These terms directly couple
the WB dynamics to LSEs in NB. For $G_{2}(\omega)$, a similar situation also occurs in
NB. Accordingly, Equations {(\ref{wbgf})} and {(\ref{nbgf})} reveal that the
non-half-integer LTQF for one orbital within OSMP plausibly arises from correlation
$\langle n_{\alpha,\sigma} \rangle$ and
$\langle d_{\alpha\downarrow}^{\dag}d_{\alpha\uparrow}\rangle$ of the other orbital.


We note that Eqs. {(\ref{wbgf})} and {(\ref{nbgf})} are obtained via a Hartree-Fock
truncation, which serves to provide an analytical and intuitive picture of the
interorbital correlation. The final numerical results presented in Figures are obtained
via the fully self-consistent DMFT without such truncation. Within the Hartree-Fock
approximation decoupling of high-order correlations, the single-particle Green’s functions
$G_{1}(\omega)$ for WB (the orbital $1$) and $G_{2}(\omega)$ for NB (the orbital $2$)
adopt an explicit form that illuminates the origin of the quasiparticle breakdown.
The schematic of this physical mechanism is shown in Fig. \ref{fig:0}. 
The equations of motion indicate that the interorbital
coupling mediated by LSEs is proportional to $J_{\rm sf}$ and also depends on $J_{\rm z}$.
When $J_{\rm sf}=J_{\rm z}=0$, NB and WB bands are decoupled in OSMP, where NB corresponds to
an insulating phase and WB to a metallic phase. When $J_{\rm sf}=J_{\rm z}\neq0$, NB and
WB become coupled in OSMP. Neither NB remains a purely insulating phase nor WB a purely
metallic phase, and the two bands exhibit strong interorbital coupling.

Fig. \ref{fig:AU} further illustrates the effect of $U$ on $A^{\rm LSE}(\omega)$. In the
metallic phase ($U = 1.0$), the LSE peaks (gray and brown dashed lines) are predominantly
located below FS with a broadly distributed lower-energy structure. In OSMP
($U=2.8, 3.0, 3.2, 3.4, 3.6$), as $U$ increases, the LSE peaks (cyan and green solid lines)
gradually shift closer to FS from initially distant positions while gaining spectral weight.
Ultimately, in the insulating phase ($U=3.8$), the LSE peaks near FS (yellow and pink dashed
lines) vanish entirely. These results indicate a clear evolution of the LSE character across
the phase diagram. In the metallic phase, the LSE spectral weight is broadly distributed,
peaking at relatively high energies. Upon entering OSMP, this weight shifts dramatically to
lower energies, condensing into sharp resonances at FS. This signifies that the energy scale
of interorbital spin excitations becomes resonant with the low-energy charge fluctuations,
enabling the strong hybridization between the bands. Ultimately, in the fully developed
Mott insulating phase, all the low-energy LSE weight is frozen out. The high concentration
of LSEs in OSMP, characteristic of a metallic state, implies that the NB (nominally insulating)
acquires partial itinerant character. This explains why LTQF in OSMP takes non-half-integer values
(i.e., $L_{o}<1/2$). Based on these results, we conclude that the LSE-driven quasiparticle
breakdown generates a partial conductive character in the nominally insulating NB and
partial localization of coherent quasiparticles in the metallic WB throughout the OSMP regime.